\documentclass[dvipsnames]{article}
\usepackage[symbol]{footmisc}

\usepackage{lineno}

\usepackage[a4paper,margin=1.1in,footskip=0.25in]{geometry}

\usepackage[utf8]{inputenc}
\usepackage[english]{babel}
\usepackage{csquotes}
\usepackage{titling}

\usepackage{graphicx}
\usepackage[usenames,dvipsnames,svgnames,table]{xcolor}

\usepackage{amssymb}
\usepackage{amsmath}
\usepackage{amsfonts}
\usepackage{amsthm}
\usepackage{algorithm}
\usepackage{algpseudocode}

\usepackage[backend=biber, style=nature, doi=false, maxbibnames=3]{biblatex}
\usepackage{xr}
\makeatletter
\newcommand*{\addFileDependency}[1]{
\typeout{(#1)}
\@addtofilelist{#1}
\IfFileExists{#1}{}{\typeout{No file #1.}}
}\makeatother

\usepackage{url}
\usepackage[hyperindex,colorlinks,hyperfootnotes = false,]{hyperref}

\newtheorem*{remark}{Remark}
\newtheorem*{notation}{Notation}
\newtheorem*{definition}{Definition}

\newcommand{\T}[0]{\mathcal{T}}

\newcommand{\x}{\mathbf{x}}
\newcommand{\X}{\mathbf{X}}

\newcommand{\E}{\mathbb{E}}

\newcommand{\R}{\mathbb{R}}
\renewcommand{\P}{\mathbb{P}}

\newcommand{\K}{\mathcal{K}}

\newcommand{\independent}{\perp\!\!\!\!\perp}

\title{Kernel-based Joint Independence Tests for Multivariate Stationary and Non-stationary Time Series}

\makeatletter
\newcommand{\printfnsymbol}[1]{
  \textsuperscript{\@fnsymbol{#1}}
}
\makeatother

\author
{Zhaolu Liu$^{1}$, Robert L. Peach$^{2,3}$ Felix Laumann$^{1}$,\\ Sara Vallejo Mengod$^{1}$ and Mauricio Barahona$^{1\ast}$
\\
\normalsize{$^{1}$Department of Mathematics, Imperial College London, London, United Kingdom}\\
\normalsize{$^{2}$Department of Neurology, University Hospital W{\"u}rzburg, W{\"u}rzburg, 97070, Germany}\\
\normalsize{$^{3}$Department of Brain Sciences, Imperial College London, London, W12 0NN, United Kingdom}, 
\\
\normalsize{$^\ast$To whom correspondence should be addressed: m.barahona@imperial.ac.uk}
}

\date{} 

\begin{document}

\maketitle

\begin{abstract}
Multivariate time series data that capture the temporal evolution of interconnected systems are ubiquitous in diverse areas. Understanding the complex relationships and potential dependencies among co-observed variables is crucial for the accurate statistical modelling and analysis of such systems. Here, we introduce kernel-based statistical tests of joint independence in multivariate time series by extending the $d$-variable Hilbert-Schmidt independence criterion (dHSIC) to encompass both stationary and non-stationary processes, thus allowing broader real-world applications. By leveraging resampling techniques tailored for both single- and multiple-realisation time series, we show how the method robustly uncovers significant higher-order dependencies in synthetic examples, including frequency mixing data and logic gates, as well as real-world climate, neuroscience, and socioeconomic data. Our method adds to the mathematical toolbox for the analysis of multivariate time series and can aid in uncovering high-order interactions in data.
\end{abstract}

\section{Introduction}
Time series that record temporal changes in sets of system variables are ubiquitous across many scientific disciplines~\cite{FULCHER2017527}, from physics and engineering~\cite{kutz_brunton_PNAS2019} to biomedicine~\cite{liu2021similarity,saavedra2021systems}, climate science~\cite{duchon2012time,katz1981use}, economics~\cite{mills1990time,siami2018forecasting} or online human behaviour~\cite{peach2019data, peach2021understanding}.
Many real-world systems are thus described 
as multivariate time series of (possibly) interlinked processes 
tracking the temporal evolution (deterministic or random) of groups of observables of interest. 
The relationships between the measured variables are often complex, in many cases displaying inter-dependencies among each other. 
For example, the spreading of Covid-19 in Indonesia was dependent on weather conditions~\cite{tosepu2020correlation};  the Sustainable Development Goals have extensive interlinkages~\cite{laumann2022complex}; there are strong interconnections between foreign exchange and cryptocurrencies~\cite{baumohl2019cryptocurrencies}; and the brain displays multiple spatial and temporal scales of functional connectivity~\cite{mokhtari2019dynamic}.
Driven by technological advances (e.g., imaging techniques in the brain sciences~\cite{guan2020profiles}, or the increased connectivity of personal devices via the Internet of Things~\cite{miorandi2012internet}), there is a rapid expansion in the collection and storage of multivariate time series data sets, which underlines the need for  mathematical tools to analyze the interdependencies within complex high-dimensional time series data.

Characterising the relationships between variables in a multivariate data set often underpins the subsequent application of statistical and machine learning methods. 
In particular, before further analyses can be performed, it is often crucial to determine whether the variables of interest are jointly independent~\cite{vogelstein2019discovering}. Joint independence of a set of $d$ variables means that no subset of the $d$ variables are dependent. 
We need not look further than ANOVA and t-tests to find classic statistical methods that assume joint independence of input variables, and the violation of this assumption can lead to incorrect conclusions~\cite{mishra2019application}. Causal discovery methods, such as structural equation modelling, also require joint independence of noise variables~\cite{pfister2018kernel}. Furthermore, joint independence has applications in uncovering higher-order networks, an emergent area highlighted in recent studies~\cite{rosas2019quantifying,battiston2021physics, arnaudon2022connecting, nurisso2023unified, battiston2020networks}.

Kernel-based methods offer a promising framework for testing statistical independence. Notably, the $d$-variable Hilbert-Schmidt independence criterion (dHSIC)~\cite{pfister2018kernel} can be used as a statistic to test the joint independence of $d$ random variables. Developed as an extension of the pairwise HSIC~\cite{gretton2007kernel} , a statistical test that measures the dependence between two variables~\cite{gretton2007kernel, chwialkowski2014kernel, chwialkowski2014wild}, dHSIC measures the dependence between $d$ variables~\cite{pfister2018kernel}.
In words, dHSIC can be simply defined as the ``squared distance'' between the joint distribution and the product of univariate marginals when they are embedded in a reproducing kernel Hilbert space (RKHS).
Crucially, kernel methods do not make assumptions about the underlying distributions or type of dependencies (i.e., they are non-parametric). Yet, in its original form, dHSIC assumes the data to be $iid$ (i.e., drawn from identical independent distributions). This is an unreasonable assumption in the case of time series data, and it has precluded its application to temporal data.

To the best of our knowledge, dHSIC has not yet been extended to time series data. The pairwise HSIC has been extended to deal with \emph{stationary} random processes under two different test resampling strategies: shifting within  time series~\cite{chwialkowski2014kernel}, and the Wild Bootstrap method~\cite{chwialkowski2014wild}. However, the assumption of stationarity, by which the statistical properties (e.g., mean, variance, autocorrelation) of the time series are assumed not to change over time, is severely restrictive in many real-world scenarios, as non-stationary processes are prevalent in many areas, e.g., stock prices under regime changes or weather data affected by seasonality or long-term trends.
Hence there is a need for independence tests that apply to both stationary and non-stationary processes.
Recently, pairwise HSIC has been extended to non-stationary random processes by using random permutations over independent realisations  of each time series, when available~\cite{laumann2021kernel}.

In this paper, we show how dHSIC can be applied to reject joint independence in the case of both stationary and non-stationary multivariate random processes. 
Following recent work~\cite{laumann2021kernel}, we adapt dHSIC so that it can be applied to stationary and non-stationary time-series data when multiple realisations are present. Additionally we develop a new bootstrap method inspired by Ref.~\cite{chwialkowski2014kernel}, which uses `shifting' to deal with stationary time-series data when only one realisation is available.
Using these methodological advances, we then introduce statistical tests that rely on these two different resampling methods to generate appropriate null distributions: one for single-realisation time series, which is only applicable to stationary random processes, and another for multiple realisation time series, which is applicable to both stationary and non-stationary random processes. 
We show numerically that the proposed statistical tests based on dHSIC identify robustly and efficiently the lack of joint independence in synthetic examples with known ground truths. We further show how recursive testing from pairwise to $d$-order joint independence can reveal emergent higher-order dependencies in real-world socio-economic time series that cannot be explained by lower order factorisations.

\section{Preliminaries}

\subsection{Kernel-based tests for joint independence}

\begin{definition}[Joint independence of a set of variables]
The $d$ variables $X^j, \, j=1,\ldots,d, \, $  with joint distribution $\mathbb{P}_{X^1, \ldots, X^d}$ are jointly independent if and only if
the joint distribution is fully factorisable into the product of its univariate marginals, i.e.,
$\mathbb{P}_{X^1, \ldots, X^d}= \prod_{j=1}^d \mathbb{P}_{X^j}$,
where the $\mathbb{P}_{X^j}$ denote the marginals.
\end{definition}

\begin{remark}[Joint independence of subsets]
If $d$ variables are jointly independent then any subset of those $d$ variables are also jointly independent, e.g.,  $\mathbb{P}_{X^1, X^2, X^3}=\mathbb{P}_{X^1} \mathbb{P}_{X^2} \mathbb{P}_{X^3}$ implies $\mathbb{P}_{X^1, X^2}=\mathbb{P}_{X^1} \mathbb{P}_{X^2}$, which follows from marginalisation with respect to $X^3$ on both sides of the equality. 
Hence, by the contrapositive, lack of joint independence of a subset of variables implies lack of joint independence of the full set of variables. 
\end{remark}

A series of papers in the last two decades has shown how kernel methods can be used to test for independence of random variables (for details see Refs.~\cite{gretton2007kernel, pfister2018kernel}).
The key idea is to embed probability distributions in
reproducing kernel Hilbert spaces (RKHS)~\cite{muandet2017kernel} via characteristic kernels, thus mapping distributions uniquely to points in a vector space. For a summary of the key definitions and foundational results see Refs.~\cite{sriperumbudur2010hilbert, sriperumbudur2011universality}.

\begin{definition}[RKHS and mean embedding for probability distributions~\cite{fukumizu2007kernel, gretton2012kernel}]
Let $\mathcal{H}_k$ be a RKHS of functions $f: \mathcal{X} \to \R$ endowed with dot product $\langle \cdot, \cdot \rangle$, and  with a reproducing kernel $k: \mathcal{X} \times \mathcal{X} \to \R$. 
Let $\P$ be a distribution defined on a measurable space $\mathcal{X}$, then the mean embedding of $\P$ in $\mathcal{H}_k$ is an element $\mu_\P \in \mathcal{H}_k$ 
given by
$\mu_\P:= \int k(\x, \cdot) \, \P(d\x)$, with the property
$\langle f, \mu_\P \rangle =  \E_\P [f] = \int f(\x) \P(d\x), \; \forall f \in \mathcal{H}_k$.
\end{definition}

If the kernel is characteristic, the RKHS mapping is injective and this representation captures uniquely the information about each distribution. 
Based on such a mapping, statistics have been constructed to test for homogeneity (using the maximum mean discrepancy, MMD~\cite{gretton2012kernel}) or independence (using the Hilbert-Schmidt independence criterion, HSIC~\cite{gretton2007kernel}) between two random variables.

\begin{remark}
An example of a characteristic kernel is the Gaussian kernel
$k_\sigma(\x, \mathbf{y}) = \exp{(-\|\x-\mathbf{y}\|^2/\sigma^2)}$ where $\x, \mathbf{y} \in \R^p$. The Gaussian kernel will be used throughout our applications below, but our results apply to any other characteristic kernel.
\end{remark}
Recently, an extension of HSIC for $d$ variables, denoted dHSIC, was introduced 
and used as a statistic for joint independence to test the null hypothesis $H_0: \P_{X^1,\cdots,X^d} = \prod_{j=1}^d\P_{X^j}$.
\begin{definition}[dHSIC~\cite{pfister2018kernel}]
Let us consider $d$ random variables $X^j, \, j=1,\ldots,d, \, $  with joint distribution $\P_{X^1, \ldots, X^d}$.
For each  $X^j$,
let $\mathcal{H}_{k^{j}}$ denote a separable RKHS with characteristic kernel $k^j$.

The $d$-variable Hilbert-Schmidt Independence Criterion (dHSIC), which measures the similarity between the joint distribution and the product of the marginals, is defined as:
\begin{align}
\label{eq:dHSIC}
    \operatorname{dHSIC}(X^1,\ldots,X^d) :=
    \left \| \mu_{\P_{X^1,\ldots,X^d}}-\mu_{\P_{X^1} \cdots \P_{X^d}}\right \|^2_{\mathcal{H}} 
\end{align}
where $ \mathcal{H} : = \mathcal{H}_{k^1}\otimes\cdots\otimes\mathcal{H}_{k^d}$ and  $\otimes$ is the tensor product. 
\end{definition}

\begin{remark}
    Given the definition~\eqref{eq:dHSIC}, 
    dHSIC is zero if and only if the variables are jointly independent, i.e., when the joint distribution is equal to the product of the marginals. This is the basis for using dHSIC to define the null hypothesis for statistical tests of joint independence.
\end{remark}

\begin{remark}[Emergent high-order dependencies]
As noted above, the rejection of joint independence for any subset of a set of $d$ variables implies also the rejection of joint independence for the full set of $d$ variables. 
Therefore, many observed rejections of joint independence at higher orders follow from rejections of joint independence at lower orders (i.e., within subsets of variables).  
To identify more meaningful high order interactions, in some cases we will also consider `first time rejections' of $d$-way joint independence, i.e., when the joint independence of a set of $d$ variables is rejected but the joint independence of each and all of its subsets of size $d'<d$ cannot be rejected. We denote these as emergent high-order dependencies.
\end{remark}

\subsection{Time series as finite samples of stochastic processes }

Our interest here is in the joint independence of time series, which we will view as finite samples of stochastic processes.

\begin{notation}[Stochastic processes and sample paths]
We will consider a set of $d$ stochastic processes $\{ X^j(t; \omega) : t \in \T \}, \,  j=1,\ldots, d$, 
where $t \in \T$ is defined over the index set, corresponding to time, and $\omega \in \Omega$ is defined over the sample space. Below, we will also use the shorthand $\{X^j_t\}$ to denote each stochastic process. 

For each stochastic process we may observe $n$ independent realisations (or paths), which are samples from $\Omega$ indexed by $\omega_i$: $\{ X^j(t; \omega_i) : t \in \T \}, \, i=1,\ldots, n$. Furthermore, each path is finite and sampled at times $t=1, \ldots, T_j$. 
%
\end{notation}

\begin{remark}[Time series as data samples]
For each variable $X^j$, the data samples (time series) consist of $n$ paths
$(X^j(1, \omega_i),\ldots, X^j(T_j,\omega_i))$, 
$i=1,\ldots, n$, 
which we arrange as $T_j$-dimensional vectors
$\x_i^j=(x^j_{i, 1}, \ldots, x^j_{i, T_j}), \, i=1,\ldots, n$, 
%
i.e., the components of the vector are given by $x^j_{i, t_k} := X^j(t_k; \omega_i)$.  
\end{remark}

\begin{definition}[Independence of stochastic processes]
Two stochastic processes $\{X^j_t\}$ and $\{X^{j'}_t\}$  
with the same index set $\T$ are independent if for every choice of sampling times $t_1, \ldots, t_f \in \T$, the random vectors $(X^j(t_{1}),\ldots ,X^j(t_{f}))$ and $(X^{j'}(t_{1}),\ldots ,X^{j'}(t_{f}))$ are independent. Independence is usually denoted as  $\{X^j_t\} \independent \{X^{j'}_t\}$. Below we will abuse notation and use the shorthand  $X^j \independent X^{j'}$.
\end{definition}

From this definition, it immediately follows that the realisations are independent.  

\begin{remark}[Independence of realisations]
Although the samples within a path $(X^j(1, \omega_i),\ldots, X^j(T_j,\omega_i)) = (x^j_{i, 1}, \ldots, x^j_{i, T_j})$  are not necessarily independent across time, each variable is independent across realisations for any time $t$, i.e.,  $x^j_{i, t} \independent x^j_{i', t} \, \forall t, \, \forall i \neq i'$.
In other words, the $n$ time series are assumed to be $iid$ samples, 
$\{\x^j_i\}_{i=1}^n  \stackrel{\text{iid}} {\sim} \, \mathbb{P}_{X^j}$, where ${P}_{X^j}$ is a finite-dimensional distribution of the stochastic process $\{X_t^j\}$. 
\end{remark}

\begin{definition} [Stationarity]
    A stochastic process is said to be stationary if all its finite-dimensional distributions are invariant under translations of time. 
\end{definition}

\begin{paragraph} 
{Aim of the paper:}
Here we use kernels to 
embed finite-dimensional distributions of the $d$ stochastic processes $\{X_t^j\}$ and design tests for joint independence of time series thereof.
Recent work has used HSIC to test for independence of pairs of stationary~\cite{besserve2013statistical,chwialkowski2014wild} and non-stationary~\cite{laumann2021kernel} time series.
Here we extend this work to $d>2$ time series using tests based on dHSIC. We consider two scenarios: 
\begin{itemize}
\item if we only observe a single time series ($n=1$) of each of the $d$ variables, then we can only consider stationary processes;
\item if we have access to several time series ($n>1$) of each of the $d$ variables, then we can also study non-stationary processes.
\end{itemize}
\end{paragraph}

\section{dHSIC for joint independence of stationary time series}
\label{sec: single}

We first consider the scenario where we only have one time series  ($n=1$) for each of the $d$ variables $X^j$, which are all assumed to be stationary.
Our data set is then
$\{\x^j\}_{j=1}^d$, and it
consists of $d$ time series vectors $\x^j=(x^j_{1}, \ldots, x^j_{T})$, which we view as single realisations of the stationary stochastic processes $\{X_t^{j}\}$, all sampled at times $t=1, \ldots, T$.
%
As will become clear below, the limited information provided by the single realisation, together with the use of permutation-based statistical tests, means that the assumption of stationarity is necessary~\cite{chwialkowski2014kernel}. 

Let $K^j \in \R^{T\times T}$ be kernel matrices with entries $K^j_{ab}=k^j(x^j_{a}, x^j_{b} )$ where $a,b \in \{1,\cdots,T\}$, and $k^j: \mathbb{R} \times \mathbb{R} \rightarrow \mathbb{R}$ is a characteristic kernel (e.g., Gaussian);
hence the matrix $K^j$ captures the autocorrelational structure of variable $X^j$. 
In this case, dHSIC~\eqref{eq:dHSIC} can be estimated as the following expansion in terms of  kernel matrices~\cite{sejdinovic2013kernel, pfister2018kernel}:
\begin{align}
 \widehat{\mathrm{dHSIC}}_{st}(\x^1, \ldots, \x^d) : = &
     \frac{1}{T^2}\sum_{a=1}^T\sum_{b=1}^T\prod_{j=1}^d K_{a b}^j - \frac{2}{T^{d+1}}\sum_{a=1}^T\prod_{j=1}^d\sum_{b=1}^T K_{a b}^j +\frac{1}{T^{2d}}\prod_{j=1}^d\sum_{a=1}^T\sum_{b=1}^T K_{a b}^j .
     \label{eq:dHSIC_single}
\end{align}
The null hypothesis is $H_0: \mathbb{P}_{X^1, \ldots, X^d}=\mathbb{P}_{X^1} \cdots \mathbb{P}_{X^d}$,
and we test~\eqref{eq:dHSIC_single} 
for statistical significance. To do so, we bootstrap the distribution under $H_0$ using random shifting to generate $S$ samples~\cite{chwialkowski2014kernel}. For each of the samples $s=1,\ldots, S$, we fix one time series ($\x^1$ without loss of generality) and generate random shifting points $\tau_{s}^j, \, j=2,\ldots,d$ for each of the other $d-1$ time series, where $ h < \tau_{s}^j < T$ and $h$ is chosen to be the first index where the autocorrelation of $\sum_{j=1}^d\x^j$ is less than 0.2~\cite{chwialkowski2014kernel}. 

\begin{figure}[htb!]
\centering\includegraphics[width=1\textwidth]{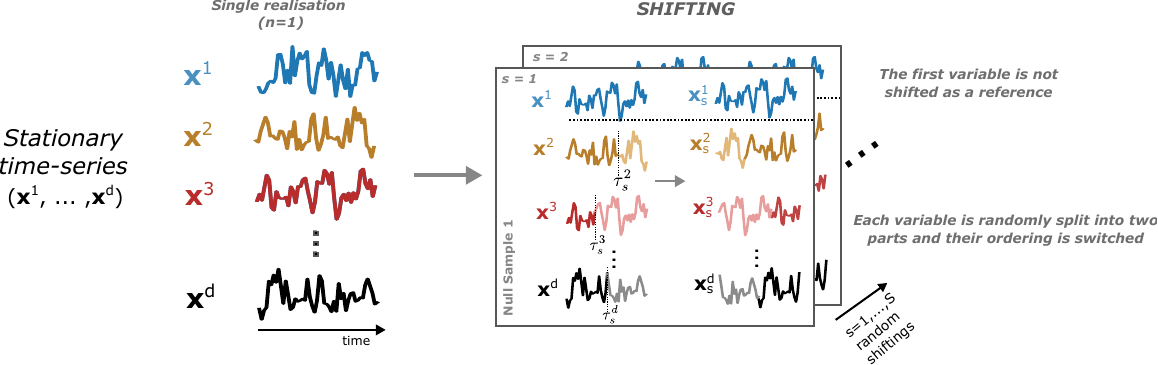}
\caption{\textbf{Shifting strategy for random sampling of single-realisation stationary time series.}
The shifting method for stationary time series when only one realisation of each variable is available. For each null sample $s$,  the first time series $\x^1$ is kept fixed and a random shifting point $\tau_j^s$ is chosen for each of the other time series $\x^j, \, j=2,\ldots,d$ so that the sections before and after $\tau_j^s$ (darker and lighter shades of colour) are switched. 
This process generates $S$ randomly shifted samples that are used to bootstrap the null distribution. 
}
\label{fig:fig1}
\end{figure}

Each time series is then shifted by $\tau_{s}^j$, so that $x_{s,t}^j 
= x_{(t+{\tau_{s}^j}) \bmod T}^j$. 
This shifting procedure, which is illustrated in Fig.~\ref{fig:fig1}, breaks the dependence across time series, yet retaining the local temporal dependence within each time series. 
In this way, we produce $S$ randomly shifted data sets
$(\x_s^1, \ldots, \x_s^d)
, s=1, \dots, S$,
and the estimated dHSIC 
is computed for each shifting:
$\widehat{\mathrm{dHSIC}}_{st}(\x_s^1, \ldots, \x_s^d)$. 
The p-value is computed by Monte Carlo approximation~\cite{pfister2018kernel}.
Given a significance level $\alpha$, 
the null hypothesis $H_0$ is rejected if $\alpha > \operatorname{p-value}$.
We note that although an alternative to shifting called Wild Bootstrap has been proposed~\cite{chwialkowski2014wild, shao2010dependent}, it has been reported to produce large false positive rates 
~\cite{mehta2019independence}. We therefore use shifting (and not the Wild Bootstrap) in this manuscript.

\subsection{Numerical results}

\subsubsection{Validation on synthetic stationary multivariate systems with a single realisation}

To validate our approach, we apply the dHSIC test for joint independence to data sets consisting of $d=3$ time series of length $T$ with $n=1$ realisations (i.e., one time series per variable). We use three stationary models with a known dependence structure (ground truth), the strength of which can be varied by 
For each test, we use $S=1000$ randomly shifted samples and we take $\alpha=0.05$ as the significance level.  
We then generate 200 such data sets for every model and combination of parameters ($T$, $\lambda$), and compute either the test power (i.e., the probability that the test correctly rejects the null hypothesis when there is dependence) or the Type I error (i.e., the probability that the test mistakenly rejects the true null hypothesis when there is independence) for the 200 data sets.

\begin{figure}[!ht]
\centering\includegraphics[width=0.65\textwidth]{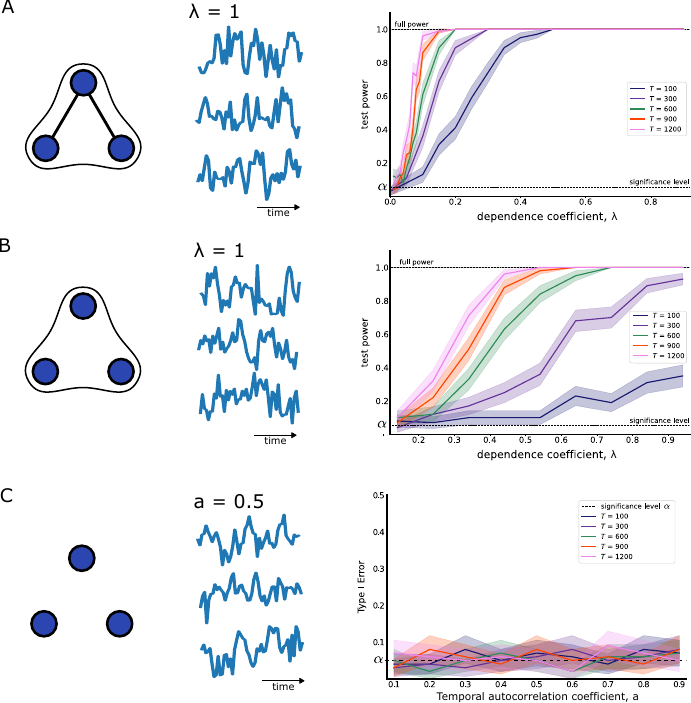}
\caption{\textbf{Stationary systems with a single realisation.} Left, a visualisation of the ground truth dependences where edges represent rejection of pairwise dependence and 3-way hyperedges represent the rejection of 3-way joint independence.
Middle, an example of the 3-variable time series for which dHSIC was computed. Right, test power (for A,B) and Type I error (for C) computed by applying the dHSIC test to 200 data sets generated for each model with different dependence coefficient $\lambda$ (autocorrelation coefficient $a$ for C) and length of time series $T$. 
The lines represent the average over the 200 data sets and the shaded area correspond to confidence intervals.
The systems are taken from Ref.~\cite{rubenstein2016kernel}: (A) 3-way dependence ensuing from pairwise dependences~\eqref{eqn:case2}; (B) Emergent 3-way dependence with no underlying pairwise dependences~\eqref{eqn:case1};  (C) Joint independence~\eqref{eqn:case3}. 
}
\label{fig:fig2}
\end{figure}

\paragraph{Model 1.1: 3-way dependence ensuing from pairwise dependences.} \label{model: 1.1}
The first stationary example~\cite{rubenstein2016kernel} has a 3-way dependence that follows from the presence of two simultaneous 2-way dependences:   
\begin{equation}
\begin{aligned}\label{eqn:case2}
X_{t} &=\frac{1}{2} X_{t-1}+\epsilon_{t},\qquad
Y_{t} &=\frac{1}{2} Y_{t-1}+\eta_{t}, \qquad
Z_{t} &=\frac{1}{2} Z_{t-1}+ \zeta_{t}+ \lambda\left(X_{t}+Y_{t}\right), 
\end{aligned}
\end{equation}
where $\epsilon_t, \eta_t, \zeta_t$ and $\theta_t$ are generated as $iid$ samples from a normal distribution $\mathcal{N}(0,1)$, and the dependence coefficient $\lambda$ regulates the magnitude of the dependence between variables, i.e., for $\lambda=0$ we have joint independence of $(X,Y,Z)$ and the dependence grows as $\lambda$ is increased. 
Figure~\ref{fig:fig2}A shows the result of our test for $d=3$ variables applied to time series of length $T=[100,300,600,900,1200]$ and increasing values of the dependence coefficient $0 \leq \lambda \leq 1 $ generated from model~\eqref{eqn:case1}.
As either $\lambda$ or $T$ increase, it becomes easier to reject the null hypothesis of joint independence. Full test power can be already reached for $\lambda=0.5$ across all lengths of time series. Our test also rejects pairwise independence between the $(X,Z)$ and $(Y,Z)$ pairs, and fails to reject independence between $(X,Y)$, as expected from the ground truth.

\paragraph{Model 1.2: Pure 3-way dependence.}
Our second stationary example, also from Ref.~\cite{rubenstein2016kernel}, includes a 3-way dependence without any underlying pairwise dependence:
\begin{equation}
\begin{aligned}\label{eqn:case1}
X_{t} &=\frac{1}{2} X_{t-1}+\epsilon_{t},\qquad
Y_{t} &=\frac{1}{2} Y_{t-1}+\eta_{t},\qquad
Z_{t} &=\frac{1}{2} Z_{t-1}+ \zeta_{t} +
\lambda\left|\theta_{t}\right| \operatorname{sign}\left(X_{t} Y_{t}\right), 
\end{aligned}
\end{equation}
where $\epsilon_t, \eta_t, \zeta_t$ and $\theta_t$ are $iid$ samples from $\mathcal{N}(0,1)$, and the coefficient $\lambda$ regulates the 3-way dependence. 
Figure~\ref{fig:fig2}B shows that the test rejects the null hypothesis as either $\lambda$ or $T$ increase, although the test power is lower relative to~\eqref{eqn:case2}, as there are no 2-way dependences present in this case, i.e., this is a 3-way emergent dependency.

\paragraph{Model 1.3: Joint independence.}
As a final validation, we use a jointly independent example~\cite{rubenstein2016kernel}:
\begin{equation}
\begin{aligned}\label{eqn:case3}
X_{t}&=a X_{t-1}+\epsilon_{t}, \qquad
Y_{t}&=a Y_{t-1}+\eta_{t}, \qquad
Z_{t}&=a Z_{t-1}+\zeta_{t} 
\end{aligned}
\end{equation}
where $\epsilon_t, \eta_t, \zeta_t$ are $iid$ samples from $\mathcal{N}(0,1)$.
Figure~\ref{fig:fig2}C shows that in this case we do not reject the null hypothesis of joint independence across a range of values of the autocorrelation parameter $a$. Note that the type I error of the test remains controlled around the significance $\alpha=0.05$ for all values of $T$ and $a$.

\subsubsection{Synthetic frequency mixing data}

\begin{figure}[!t]
\centering\includegraphics[width=1\textwidth]{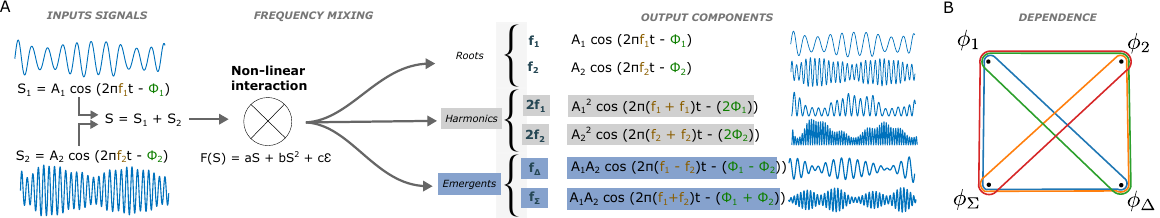}
\caption{\textbf{Frequency mixing.} (A) Two independent input signals with root frequencies $f_1=$7 Hz and $f_2=$18 Hz are mixed via a quadratic function with noise to generate the signal $F$. This signal has components at the root frequencies, harmonics (double of the root frequencies), and emergents (sum and difference of the root frequencies), as shown by the output waveforms (right). Time series of the instantaneous phases $\phi_1, \phi_2, \phi_\Delta, \phi_\Sigma$ are extracted from the output components at $f_1, f_2, f_\Delta, f_\Sigma$ and dHSIC is applied to them.
(B) In this case, the dHSIC test does not reject pairwise independence between any pair of variables (i.e. there are no pairwise dependencies), but rejects the joint independence between any three of the four variables (i.e. four 3-way emergent dependencies are present), as shown by the triangles, and, consequently, also rejects the joint independence between the four variables (square).} 
\label{fig:fig3}
\end{figure}

As a further illustration linked more closely to real-world applications, we have generated a data set based on frequency mixing of temporal signals. Frequency mixing is a well known phenomenon in electrical engineering, widely used for heterodyning, i.e., shifting signals from one frequency range to another. Applying a non-linear function (e.g., a quadratic function or a rectifier) to the sum of two signals with distinct frequencies generates new signals with emergent frequencies at the sum and difference of the input signals (Figure~\ref{fig:fig3}A-C). It has previously been shown that the instantaneous phases of the emergents display a unique 3-way dependence, without any pairwise dependences~\cite{luff2023neuron,haufler2019detection,kleinfeld2006spectral}. Importantly, given sufficiently long time series 
the instantaneous phase can be considered a stationary signal~\cite{luff2023neuron}. Hence we can apply our test to this system. 

Here, we generated a data set using the sum of two sinusoidal functions with frequencies $f_1=$7Hz and $f_2=$18Hz as input, to which we applied a quadratic function plus weighted Gaussian noise $\epsilon$. This produces a signal $F$ that contains components at input ('root') frequencies ($f_1=$7Hz and $f_2=$18Hz), second harmonics ($2 f_1=$14Hz and $2 f_2=$36Hz), and emergent frequencies ($f_\Delta= f_2-f_1=$11Hz and $f_\Sigma=f_1+f_2=$25Hz). See Figure~\ref{fig:fig3}A and Ref.~\cite{luff2023neuron} for further details. We then computed a wavelet transform and extracted the instantaneous phases for frequencies $f_1, f_2, f_\Sigma$ and $f_\Delta$, which we denoted $\phi_1, \phi_2, \phi_\Sigma$ and $\phi_\Delta$. These phases can be considered as stationary time series.   
The ground truth is that there should be no pairwise dependencies between any of those phases, but there are higher order interactions involving 3-way and 4-way dependencies~\cite{luff2023neuron}. 

We applied dHSIC with shifting to all possible groupings of $d$ phases (for $d=2,3,4$) from the set $\{\phi_1, \phi_2, \phi_\Sigma, \phi_\Delta\}$. The phases consisted of time series with length $T=1000$, and we used $S=1000$ shiftings for our bootstrap. We found that the null hypothesis of independence could not be rejected for any of the six phase pairs ($d=2$), whereas joint independence was rejected for all four phase triplets ($d=3$) and for the phase quadruplet ($d=4$).  
The rejection of all the 3-way and 4-way joint independence hypotheses, without rejection of any of the pairwise independence hypotheses, thus recovers the ground truth expected structure (Figure~\ref{fig:fig3}B).

\subsubsection{Application to climate data}

As an application to real-world data, we used the PM2.5 air quality data set, which contains four variables: hourly measurements of the Particular Matter 2.5 (PM2.5) recorded by the US Embassy in Beijing between 2010 and 2014, and three concurrent meteorological variables (dew point, 
temperature, air pressure) 
measured at Beijing Capital International Airport~\cite{liang2015assessing}. Non-stationary trends and yearly seasonal effects were removed by taking differences of period 1 and period 52 in the averaged weekly data. Stationarity of the de-trended series was verified by an Adfuller test~\cite{said1984testing}. 
As expected, we find that the null hypotheses (joint independence) are rejected for all groups of $d=2,3,4$ variables, implying that PM2.5, dew point, temperature and air pressure are all dependent on each other.

\section{dHSIC for joint independence of non-stationary time series with multiple realisations}
\label{sec: multi}

When we have multiple independent observations of the $d$ variables, these can be viewed as $iid$ samples of a multivariate probability distribution. By doing so, the requirements of stationarity and same point-in-time measurements across all variables can be loosened.

Consider the case when we have access to $n>1$ observations of the set of variables $(X^1, \ldots, X^d)$, where each observation $i=1,\ldots, n$ consists of  $d$ time series $X^j$, which we write as vectors $\x_i^j= (x^j_{i,1}, \ldots, x^j_{i,T_j})$ of length $T_j$. Each of the $n$ observations thus consists of a set 
$\{\x^j_i\}_{j=1}^d$,
which can be viewed as an independent ($iid$) realisation of a finite-dimensional multivariate distribution $\mathbb{P}_{X^1, \ldots, X^d}$.
To simplify our notation, we compile the $n$ observations of each $X^j$ as rows of a $n \times T_j$ matrix $\X^j$, so that $\X^j_{[i,:]}=\x^j_i$.

\begin{figure}[!ht]
\centering\includegraphics[width=1\textwidth]{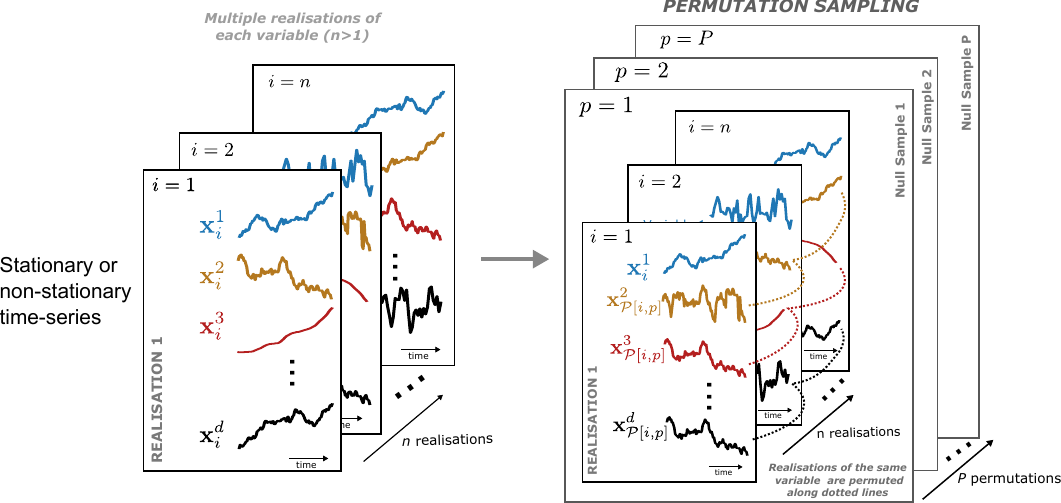}
\caption{\textbf{Random permutation sampling of multivariate time series with multiple realisations.}
A permutation strategy similar to the one developed for $iid$ data~\cite{pfister2018kernel} 
can be applied when multiple realisations of either stationary or non-stationary time series data are available. Each null sample $p=1,\ldots, P$ is generated by randomly permuting the time series for variables $j=2,\ldots, d$ across realisations $i=1,\dots, n$, as indicated by the dotted lines, whilst the first variable remains unchanged. Null distributions are generated from independent samples of this process.}
\label{fig:fig1b}
\end{figure}

Let $\kappa^j: \mathbb{R}^{T_j} \times \mathbb{R}^{T_j}\rightarrow \mathbb{R}$ be a characteristic kernel (e.g., Gaussian) that captures the similarity between a pair of time series of variable $X^j$. 
We then define the set of kernel matrices $\K^j \in \R^{n\times n}$ with entries $\K^j_{\alpha \beta}=\kappa^j(\x^j_{\alpha}, \x^j_{\beta} )$ where $\alpha,\beta \in \{1,\cdots,n\}$.
Therefore the matrix $\K^j$ captures the similarity structure between the time series of variable $X^j$ across the $n$ observations. This setup thus allows us not to require stationarity in our variables, since the $n$ observations capture the temporal behaviour of the $d$ variables concurrently.
In this case, dHSIC for the set of observations $(\X^1, \ldots, \X^d )$ can be estimated as~\cite{pfister2018kernel}:
\begin{align}
\label{eq:dHSIC_mult}
     \widehat{\mathrm{dHSIC}}_{mult}(\X^1, \ldots, \X^d):=&
     \frac{1}{n^2}\sum_{\alpha=1}^n\sum_{\beta=1}^n\prod_{j=1}^d \K^j_{\alpha \beta}
     - \frac{2}{n^{d+1}}\sum_{\alpha=1}^n\prod_{j=1}^d\sum_{\beta=1}^n \K^j_{\alpha \beta}
    +\frac{1}{n^{2d}}\prod_{j=1}^d\sum_{\alpha=1}^n\sum_{\beta=1}^n \K^j_{\alpha \beta}.
\end{align}
Similarly to Section~\ref{sec: single}, the null hypothesis is $H_0: \mathbb{P}_{X^1, \ldots, X^d}=\mathbb{P}_{X^1} \cdots \mathbb{P}_{X^d}$ and we test~\eqref{eq:dHSIC_mult} 
for statistical significance. Due to the availability of multiple realisations, however, we use a different resampling method (standard permutation test) to bootstrap the distribution of~\eqref{eq:dHSIC_mult} under $H_0$ (Fig.~\ref{fig:fig1b}). For each of the samples $p=1, \ldots, P$, we fix one variable ($X^1$ without loss of generality), and we randomly permute the rest of the variables \emph{across realisations} to create the permuted sample 
 $\{\x^j_p\}_{j=1}^d= \{\x^j_{\mathcal{P}[i,p]}\}_{j=1}^d$, where $\mathcal{P}[i,p]$ indicates a random permutation between realisations, and $\x^1_p=\x^1_i,\, \forall p$.
In this way, we produce $P$ permuted data sets
$(\X_p^1, \ldots, \X_p^d ),  \, p=1, \dots, P$, with $\X_p^1=\X^1$.
The estimated dHSIC~\eqref{eq:dHSIC_mult}
is then computed for each permutation $p$.
Given a significance level $\alpha$, the null hypothesis $H_0$ is rejected if $\alpha > \operatorname{p-value}$ where the p-value is computed by Monte Carlo approximation~\cite{pfister2018kernel}.

\subsection{Numerical results}

\subsubsection{Validation on simple non-stationary multivariate systems}

\begin{figure}[htb]
\centering\includegraphics[width=1\textwidth]{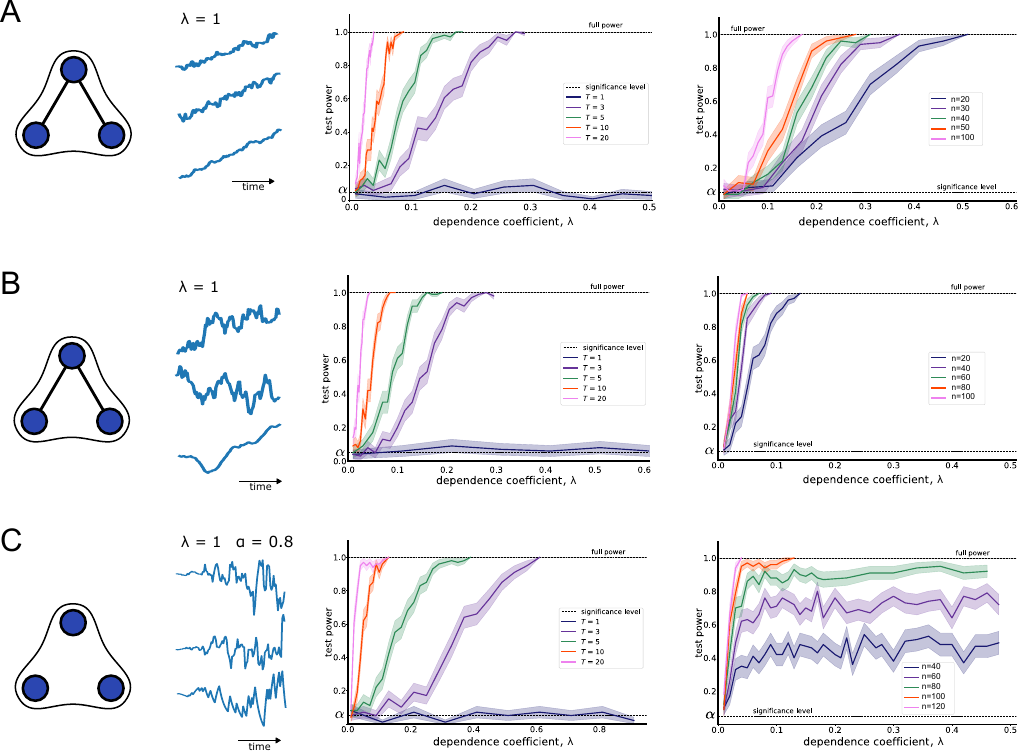}
\caption{\textbf{Three-variable non-stationary systems with multiple realisations.} Left, a visualisation of the ground truth dependences: edges represent pairwise dependence, 3-edges represent a 3-way dependence.
Middle left, an example of a realisation of a 3-variable time series. Middle right, test power computed by applying the dHSIC test to 200 data sets generated from a model at varying values of the dependence coefficient $\lambda$ and the length of time series $T$, with fixed number of realisations $n=100$.
Right, test power computed by applying the dHSIC test to 200 data sets generated from a model at varying values of the dependence coefficient $\lambda$ and the number of realisations $n$, with fixed length of time series $T=20$.
The lines represent the average over the 200 data sets and the shaded areas correspond to confidence intervals. The systems are:
(A)-(B) 3-way dependence ensuing from 2-way dependences~\eqref{eq:model_2_1}: (A) linear trend, (B) a complex dependence term; (C) 3-way dependence with no underlying pairwise dependences and a non-stationary trend~\eqref{eq:model_2_2}. 
}
\label{fig:fig4}
\end{figure}

The dHSIC test is applied to data sets consisting of $n$ observations of non-stationary time series of length $T$ of three variables $(X,Y,Z)$, with ground truth dependences that can be made stronger by increasing a dependence coefficient $\lambda$. For every model and combination of parameters ($n$, $T$, $\lambda$), we generate 200 data sets and compute the test power, i.e., the probability that the test correctly rejects the null hypothesis in our 200 data sets. 
Figure~\ref{fig:fig4} shows our numerical results for two non-stationary models: the first model (shown in Fig.~\ref{fig:fig4}A-B with two non-stationary trends) has a 3-way dependence ensuing from 2-way dependences; the second model (shown in Figure~\ref{fig:fig4}D for a non-stationary trend) has an emergent 3-way dependence with no pairwise dependences. 

\paragraph{Model 2.1: 3-way dependence ensuing from pairwise dependences with non-stationarity. }
The first model has the same dependence structure as~\eqref{eqn:case2}, i.e., two simultaneous pairwise dependences and an ensuing 3-way dependence, but in this case with non-stationary trends: 
\begin{equation}
\begin{aligned}
    X_{t} = g_1(t) + X_{t-1}+ \epsilon_{t},\qquad
    Y_{t} = g_2(t) + Y_{t-1} + \eta_{t},\qquad
    Z_{t} = g_3(t) + Z_{t-1} + \zeta_{t} + \lambda\left(X_{t}+Y_{t}\right)
    \label{eq:model_2_1}
\end{aligned}
\end{equation}
where $\epsilon_t, \eta_t, \zeta_t$ are $iid$ samples from a normal distribution $\mathcal{N}(0,1)$; $\lambda$ regulates the strength of the dependence ($\lambda=0$ means joint independence); and $g_1(t), g_2(t), g_3(t)$ are non-stationary trends:
\begin{itemize}
    \item Linear trend (Fig.~\ref{fig:fig4}A): 
    $g_1(t)=g_2(t)=g_3(t)=t$
    \item Complex non-linear trend (Fig.~\ref{fig:fig4}B): $g_1(t)=\frac{\sin^2 (t)}{\log(1+t)}, \quad  g_2(t)=\frac{\cos^2 (t)}{\log(1+t)}, \quad  g_3(t)= \frac{\sin(t)\cos(t)}{\log(1+t)}$.   
\end{itemize}
%
Figure~\ref{fig:fig4}A-B shows that the dHSIC test is able to reject the null hypothesis of joint independence for~\eqref{eq:model_2_1} even for short time series and small values of the dependence coefficient $\lambda$. The test power increases rapidly as the length of the time series $T$ or the number of realisations $n$ are increased.  As expected, the null hypothesis cannot be rejected for $T=1$, since the temporal dependence is no longer observable. 

\paragraph{Model 2.2: Emergent 3-way dependence with non-stationarity.}
The second model has the same dependence structure as~\eqref{eqn:case3} (i.e., an emergent 3-way dependence without 2-way dependences) but with non-stationary trends: 
\begin{equation}
\begin{aligned}
    X_{t} = a X_{t-1} + \epsilon_{t} +t \sin (t), \qquad
    Y_{t} = a Y_{t-1} + \eta_{t} +t \cos (t), \qquad
    Z_{t} = a Z_{t-1}+ \zeta_{t} + \lambda \, t \operatorname{sign}(X_{t} Y_{t})
    \label{eq:model_2_2}
\end{aligned}
\end{equation}
where, again, $\epsilon_t, \eta_t, \zeta_t$ are $iid$ samples from $\mathcal{N}(0,1)$, and $\lambda$ regulates the strength of the dependence. We set $a=0.8$, the point at which the data becomes non-stationary according to an Adfuller test. 
 Fig.~\ref{fig:fig4}C shows good performance of the test, which is able to reject joint independence for low values of $\lambda$, with increasing test power as the length of the time series $T$ and the number of realisations $n$ is increased (Fig.~\ref{fig:fig4}C).

\subsubsection{Synthetic XOR dependence}
The Exclusive OR (XOR) gate (denoted $\oplus$) is a logical device with two Boolean (0-1) inputs and one Boolean output, which returns a 1 when the number of `1' inputs is odd.
Here we consider a system with 3 Boolean variables $X,Y,W$ driven by noise, which get combined via XOR gates to generate another Boolean variable $Z$:
\begin{equation}
    \begin{aligned}
        X_t &= \begin{cases}
                X_{t-1},& \text{if } \epsilon_t\geq 0.5\\
                1-X_{t-1}, & \text{otherwise}
                \end{cases} \qquad\qquad
        Y_t = \begin{cases}
                Y_{t-1},& \text{if } \eta_t\geq 0.5\\
                1-Y_{t-1}, & \text{otherwise}
                \end{cases}\\
       W_t &= \begin{cases}
                W_{t-1},& \text{if } \zeta_t\geq 0.5\\
                1-W_{t-1}, & \text{otherwise}
                \end{cases} \qquad\qquad
        Z_t = \begin{cases}
                X_{t}\oplus Y_{t} \oplus W_{t},& \text{if } \theta_t\geq 0.5\\
                1-X_{t}\oplus Y_{t} \oplus W_{t}, & \text{otherwise}
                \end{cases}
    \end{aligned}
    \label{eq:XOR_example}
\end{equation}
where $\epsilon_t, \eta_t, \zeta_t, \theta_t$ are $iid$ samples from $\mathcal{U}[0,1)$, a uniform distribution between 0 and 1, and $X_0, Y_0, W_0$ are initialised as random Boolean variables.
The dependence in this system is high-order: it only appears when considering the 4-variables, with no 3-way or 2-way dependences. We find that our test does not reject joint independence for $d=[2,3]$ variables, but does reject joint independence of the 4-variable case.

\subsubsection{Application to MRI and Alzheimer's data}
\label{sec:dementia}
As a first application to data with multiple realisations, we apply our test to the MRI and Alzheimer's longitudinal data set~\cite{mri_data}, which comprises demographic and Magnetic Resonance Imaging (MRI) data collected from subjects over several visits. Here, we consider $n=56$ subjects, each with at least 3 visits ($T=3$), and we assume that the subjects constitute $iid$ realisations, a reasonable assumption since this is a well-designed population study with representative samples. We then perform dHSIC tests to find dependencies between four key variables: Age, Normalised Whole Brain Volume (nWBV), Estimated total intracranial volume (eTIV) and Clinical Dementia Rating (CDR). The first three variables are clinical risk factors, whereas CDR is a standardised measure of disease progression.

\begin{figure}[!htb]
\centering\includegraphics[width=0.85\textwidth]{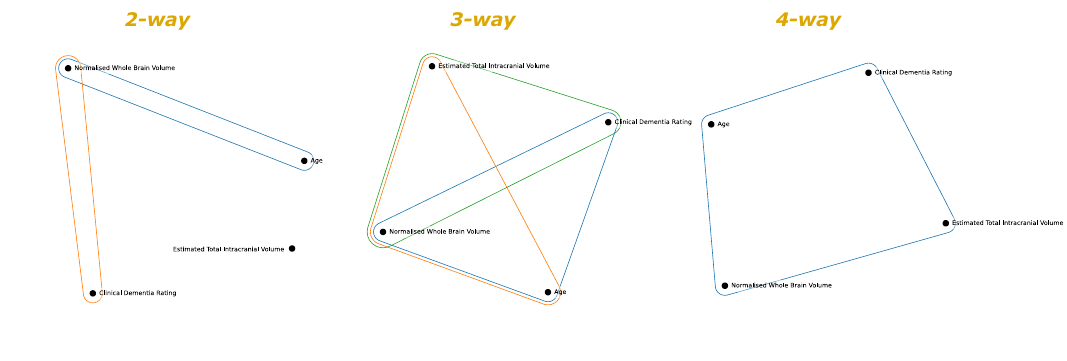}
\caption{\textbf{High-order dependencies between four variables in the MRI and Alzheimer's data containing multiple realisations of time series data.} The hyperedges represent rejections of the respective joint independence tests.
We find 2 (out of 6) pairwise dependences, 3 (out of 4) 3-way dependences, as well as the 4-way dependence between all variables. There are no emergent dependencies in this example.
}
\label{fig:dementia}
\end{figure}

Our findings are displayed as hypergraphs in Figure~\ref{fig:dementia} where nodes represent variables and hyperedges represent rejections of joint independence from the 2-way, 3-way and 4-way dHSIC tests. 
In this case, we find only 2 pairwise dependencies (Age-nWBV and nWBV-CDR) whilst eTIV is seemingly disconnected to the rest of the variables. Note that the possible emergent 3-way interaction (Age-eTIV-CDR) is not present, although eTIV shows the expected 3-way and 4-way dependences with CDR, nWBV and Age. 
This example highlights how our method can be used to reveal the different higher-order dependencies beyond pairwise interactions. To understand the complex high-order interactions of the incomplete factorisations, methods based on Streitberg and Lancaster interaction can be explored in future work~\cite{liu2023interaction}.

\subsubsection{Application to socioeconomic data}\label{sec: SDG}

As a final illustration in a different domain area, we test for joint independence between the United Nations’ Sustainable Development Goals (SDGs)~\cite{worldbank}. This data set consists of time series of a large number of socioeconomic indicators conforming the 17 SDGs (our variables $X^j, \, j=1, \ldots, 17$) measured yearly between 2000 and 2019 ($T=20$) for all 186 countries in the world (see Ref.~\cite{laumann2022complex} for details on the data set). 
We take the countries to be $iid$ realisations, as in Ref.~\cite{laumann2022complex}, although this assumption is less warranted here than for the dementia data set in Section~\ref{sec:dementia}, due to moderate correlations between countries due to socio-economic and political relationships. 
As an illustration of the differences in data dependences across country groupings, we consider two classic splits: (i) a split based on income level ($n=74$ countries with low and lower middle income and $n=105$ countries with high and upper middle income); (ii) a split based on broad geography and socio-economic development ($n=49$ countries in the Global North and $n=137$ countries in the Global South).
This data set highlights the difficulties of examining high-order dependences as the number of variables grows, e.g., $d=17$ in this case. 

\begin{figure}[!ht]
\centering\includegraphics[width=1\textwidth]{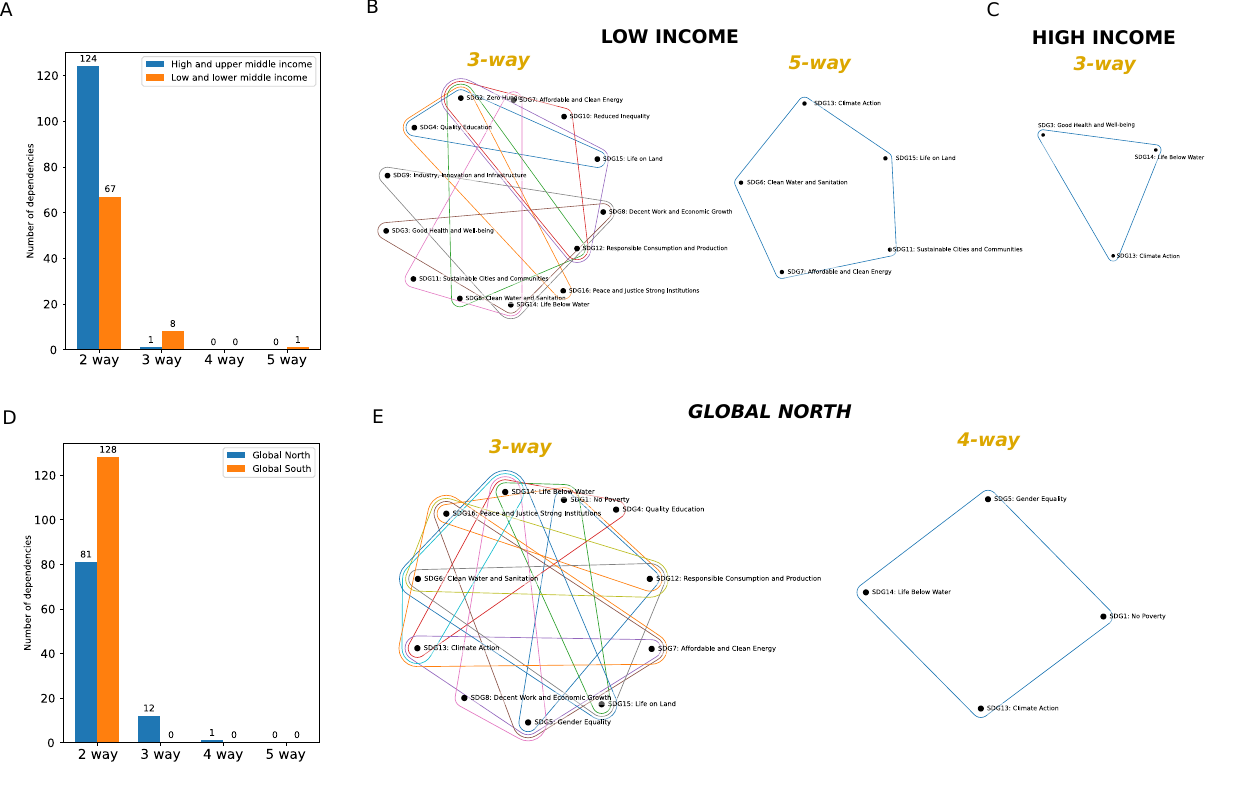}
\caption{\textbf{Emergent high-order dependences between Sustainable Development Goals (SDGs).} (A-C) Comparison of SDG dependences in low and high income countries. (A) There is a higher number of emergent higher-order dependences in low-income countries. The $d > 2$ dependences are mapped onto $d$-order hypergraphs for (B) low-income and (C) high-income countries. (D-E) Comparison of SDG dependences in Global North and Global South countries. (D) Emergent high-order dependences are found in the Global North, whereas the Global South displays only 2-way dependences.  (E) The $d > 2$ dependences for the Global North are mapped onto $d$-order hypergraphs. 
}
\label{fig:fig5}
\end{figure}

The results of applying this recursive scheme to the SDG data set are shown in Figure~\ref{fig:fig5}. The comparison between lower and high income countries (Figs.~\ref{fig:fig5}A-C) shows that higher income countries have strong pairwise dependences (124 rejections of 2-way independence out of a total of 136 pairs) and only 1 emergent 3-way interaction (Fig.~\ref{fig:fig5}C), whereas lower income countries have more emergent higher-order dependences (eight 3-way and one 5-way) (Fig.~\ref{fig:fig5}B). These results suggest that the interdependences between SDGs are more complex for lower income countries, 
whereas most of the high order dependences in high-income countries are explained away by the pairwise dependences between indicators. Given that many analyses of SDG interlinkages consider only pairwise relationships, this implies the need to consider high-order interactions to capture relationships in lower income countries where policy actions targeting pairwise interlinkages could be less effective.
The comparison between the Global North and Global South (Figs.~\ref{fig:fig5}D-E) shows that the Global South has exclusively 2-way dependences, whereas the Global North has emergent 3-way interactions (12) and 4-way (1) (Fig.~\ref{fig:fig5}E).
Interestingly, two SDGs, climate action and life below water, consistently appear in emergent high-order dependences in lower income, higher income, and in Global North groupings, suggesting their potential for further studies. In addition, the hypergraphs of emergent high-order interactions for different country groupings can be studied using network science techniques, including the computation of centrality measures to rank the importance of SDGs within the system of interdependent SDG objectives, and the use of  community detection algorithms to extract clusters of highly interdependent SDGs~\cite{laumann2022complex}.

\section{Discussion}

In this paper, we present dHSIC tests for joint independence in both stationary and non-stationary time series data. For single realisations of stationary time series, we employ a random shifting method as a resampling technique. In the case of multiple realisations of either stationary or non-stationary time series, we consider each realisation as an independent sample from a multivariate probability distribution, enabling us to utilise random permutation as a resampling strategy.
To validate our approach, we conducted experiments on diverse synthetic examples, successfully recovering ground truth relationships, including in the presence of a variety of non-stationary behaviours. 
As illustrated by applications to climate, Sustainable Development Goal and the MRI and Alzheimer's data, the testing framework could be applicable to diverse scientific areas in which stationary or non-stationary time series are the norm.

\begin{figure}[ht!]
\centering\includegraphics[width=1\textwidth]{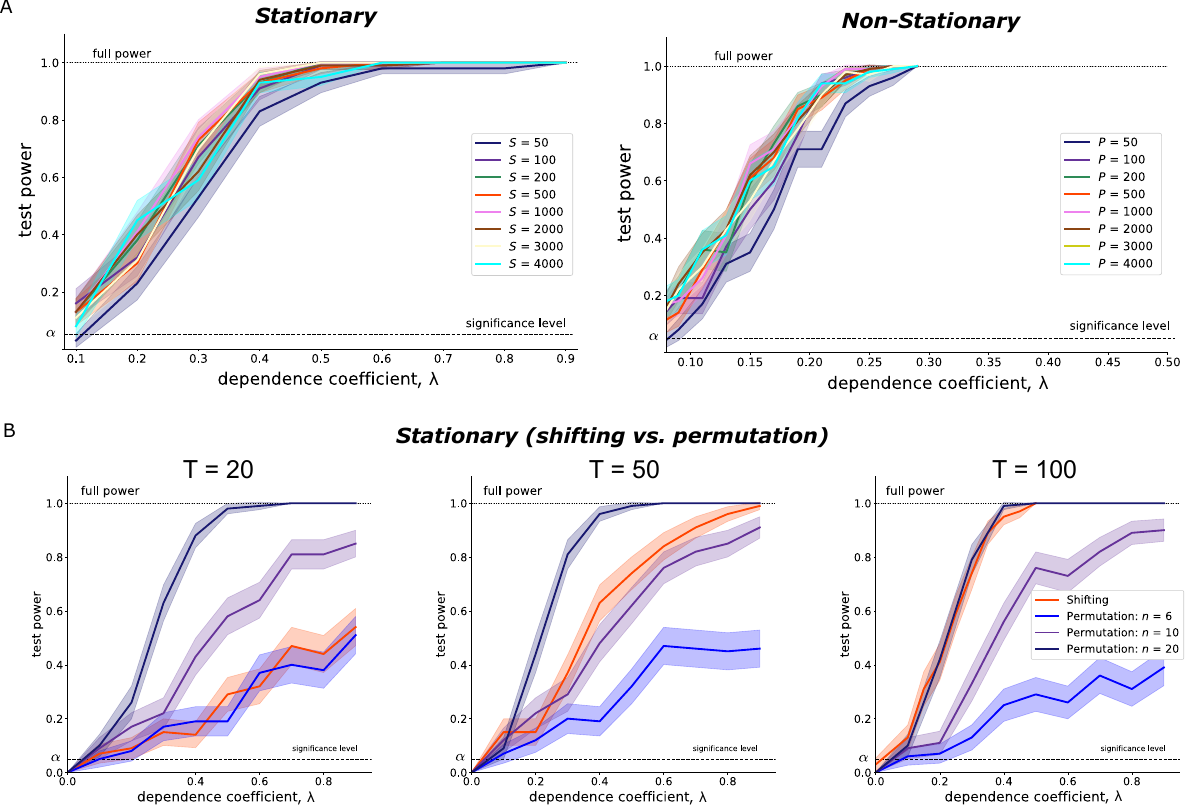}
\caption{\textbf{Robustness and efficiency of shifting and permutation resampling strategies.}
(A) Test power for the same model in its stationary version (Model 1.1~\eqref{eqn:case2}, using shifting resampling, left) and non-stationary version (Model 2.1~\eqref{eq:model_2_1} with linear trend, using permutation resampling, right). Relatively few null samples ($S,P > 100$) are enough to attain high test power for both schemes. (B) For stationary time-series with multiple realisations, both shifting and permutation can be employed. Shifting is preferred if the number of time observations ($T$) is large relative to the number of realisations ($n$); conversely, permutation is preferred if $n$ is large relative to $T$. For Model 1.1~\eqref{eqn:case2} with $T=20$, the permutation scheme with $n=6$ already reaches comparable performance to shifting, whereas for $T=100$ we need $n=20$ for permutation resampling to reach comparable performance to shifting. We use $S=P=1000$ for all tests in (B).
}
\label{fig:fig6}
\end{figure}

There are some computational considerations that need to be taken into account for different applications.
In our numerical experiments, we have evaluated the impact of several parameters, including the length of the time series $T$ and the number of observations $n$, on the computational efficiency and statistical power of our test.
In general, the test statistic can be computed in $\mathcal{O}(d T^2)$ or $\mathcal{O}(d n^2)$, where $d$ is the number of variables and $T^2$ or $n^2$ are the sizes of the kernel matrices~\cite{pfister2018kernel}. Hence the computational cost increases with the number of variables and/or number of realisations and length of time series.
The computational cost also grows linearly with the number of resamplings ($S$ or $P$) used to approximate the null distribution, but our findings show that the test is robust even for low numbers of resamplings.  
Figure~\ref{fig:fig6}A shows that the test power does not improve substantially beyond 100 resamplings (permutations), a result that has been previously discussed for $iid$ data~\cite{rindt2021consistency}
Therefore achieving a balance between test power and computational efficiency is crucial, particularly when dealing with large multivariate data sets.

It is worth noting that for stationary data with multiple independent realisations, both resampling schemes (shifting and permutation) can be employed to sample the null distribution. 
If the number of realisations ($n$) is much larger than the length of the time series ($T$), the permutation strategy provides more efficient randomisation as long as the realisations are diverse.
Conversely, when $n$ is smaller than $T$, time shifting allows to exploit better the observed temporal dynamics.
As an illustration of this point for Model 1.1~\eqref{eqn:case2} with multiple realisations,  Figure~\ref{fig:fig6}B shows that if we have $T=20$ time points available, then the permutation-based approach has the same performance as the shifting approach when the number of realisations reaches $n=6$. However, if $T=100$ time points are available, both performances become similar when the number of realisations is $n=20$. 
These resampling  alternatives must be also evaluated in conjunction with the study of different kernels that can more effectively capture the temporal structure within and across time series (e.g., signature kernels).  
We leave the investigation of these areas as an avenue of future research.

The interest in higher-order networks, such as hypergraphs or simplicial complexes, has been steadily growing~\cite{battiston2020networks} with applications across scientific fields~\cite{gao2018studying, arenzon1993neural, sonntag2004competition, klamt2009hypergraphs, arnaudon2022connecting}. Higher-order networks can be natural formalisations of relational data linking $d$ entities~\cite{white1986structure, atkin1974mathematical}. However, there is a scarcity of research and a lack of consensus on how to \emph{construct} higher-order networks from observed $iid$ or time series data~\cite{schneidman2003network}, and the joint independence methods proposed here could serve to complement approaches based on information measures~\cite{rosas2019quantifying}. 
By iteratively testing from pairwise independence up to $d$-order joint independence, our approach can uncover emergent dependencies not explained by lower-order relationships. This framework presents a direction for the development of higher-order networks, bridging the gap between observed data and the construction of meaningful higher-order network representations.

\section*{Ethics}
Ethics approval was not required for this study

\section*{Data Access}
Climate data: \url{http://dx.doi.org/10.24432/C5JS49}, 
SDG data: \url{ https://datacatalog.worldbank.org/dataset/sustainable-development-goals}, 
and MRI and Alzheimer's data: \url{https://www.kaggle.com/datasets/jboysen/mri-and-alzheimers}.
Synthetic examples: The code to generate the synthetic data and the algorithms to implement the tests are available at: \url{https://github.com/barahona-research-group/dHSIC_ts}.

\section*{Author Contribution}
ZL, RP, and MB conceived the study. ZL conducted all kernel-based tests, supervised by RP and MB. RP created the frequency mixing data. FL and SM prepared the SDG data. ZL, RP, FL, SM and MB drafted the manuscript. All authors reviewed the manuscript before submission.

\section*{Competing Interest}
The authors declare no competing interests.

\section*{Acknowledgement}
MB acknowledges support by EPSRC grant EP/N014529/1 funding the EPSRC Centre for Mathematics of Precision Healthcare at Imperial, and by the Nuffield Foundation under the project ``The Future of Work and Well-being: The Pissarides Review". RP acknowledges funding from the Deutsche Forschungsgemeinschaft (DFG, German Research Foundation) Project-ID 424778381-TRR 295.

We thank Jianxiong Sun for valuable discussions, Asem Alaa for help in maintaining the GitHub repository.

\printbibliography
\end{document}